\documentclass{INTERSPEECH2023}
\usepackage{multirow}
\usepackage{amsmath}

\interspeechcameraready

\title{ChinaTelecom System Description to VoxCeleb Speaker Recognition Challenge 2023}
\name{Mengjie Du, Xiang Fang, Jie Li}
\address{
  China Telecom Corporation Ltd. Data\&AI Technology Company, Beijing, China}
\email{\{dumj2, fangx1, lij86\}@chinatelecom.cn}

\begin{document}

\maketitle
 
\begin{abstract}
This technical report describes ChinaTelecom system for Track 1 (closed) of the VoxCeleb2023 Speaker Recognition Challenge (VoxSRC 2023).
Our system consists of several ResNet variants trained only on VoxCeleb2, which were fused for better performance later. 
Score calibration was also applied for each variant and the fused system. 
The final submission achieved \textit{minDCF} of 0.1066 and \textit{EER} of 1.980\%.
\end{abstract}

\section{Fully Supervised Speaker Verification}

The track 1 of VoxSRC 2023 in this year is a fully supervised speaker verification task, where participants can use only VoxCeleb2 development set \cite{chung2018voxceleb2} for training. For this track, we trained 7 ResNet variants with varying sizes. Details are as followed.

\subsection{Data augmentation}

We applied the similar data augmentation method from the SJTU online system \cite{chen2022sjtu}. 
\begin{itemize}
  \item Do speed perturbation with the ratio from {0.9, 1.0, 1.1} randomly, expanding the number speaker classes by a factor of 3.
  \item Randomly decide whether to do noise augmentation with the ratio of 0.6. If do, randomly select noise from MUSAN \cite{snyder2015musan} and RIR \cite{ko2017study} datasets.
  \item Randomly select a fixed length segment from the current utterance.
\end{itemize} 
For all models, 80-dimension Mel filter banks (fbank) of a 2s segment were taken as input, with a 25ms frame length and 10ms frame shift. 

\subsection{Model architectures}

ResNet has achieved state-of-the-art performance in speaker recognition within recent years.
Thus, we chose ResNet and its variant, the Res2Net-based architectures, as the foundational backbone networks.
Specifically, we incorporated Res2Net \cite{gao2019res2net} and ERes2Net \cite{Chen2023eres2net} to leverage both local and global speaker information for enhanced discriminative speaker embeddings. 
To elevate the complexity of architectures, we expanded the depths of these three frameworks (ResNet, Res2Net, ERes2Net) to 152 and 293, respectively.
Furthermore, We also integrated SimAm-ResNet293 \cite{qin2022simple} as an auxiliary model for performance improvement.

\subsection{Pooling layer}

Pooling methods coupled with attention mechanisms have been proven effective in speaker verification, which assign larger weights to more discriminative speaker characteristics. 
In this case, we employed multi-query multi-head attention pooling method (MQMHA) \cite{Zhao2022mqmha} to alleviate model sticking in some certain patterns. 
We set the head number to 8, the query number to 2, the scale factor to 2, and the final speaker embedding dimension to 256.


\subsection{Loss functions}

We employed AAM \cite{deng2019arcface}, K-subcenter \cite{deng2020sub} with Inter-TopK \cite{Zhao2022mqmha} as loss function to train all single backbone networks, 
where the scale and margin were set to 32 and 0.2 in AAM loss, the sub-center number K was set to 3, and the penalty and TopK were set to 0.06 and 5, respectively.

We employed Sphereface2 \cite{wen2021sphereface2,Han2023sphere} as loss function for models with depth of 293. The weight of positive and negative pairs was set to 0.7, and the margin was set to 0.2 working as C-type.
\begin{table}[!htbp]
  \centering
  \caption{Network architectures of system. The base channel number of all models is 32.}
  \resizebox{1.\columnwidth}{!}{
  \begin{tabular}{ccccc}
  \hline
  \multirow{2}{*}{Network}         & \multirow{2}{*}{ID} & \multicolumn{2}{c}{Step 1}                            & Step 2                    \\ \cline{3-4}
                                   &                     &   AAM+K-Subcenter+Inter-TopK                    & Sphereface2               & AAM+K-subcenter           \\ \hline
  ResNet152                        & 1                   & \checkmark &                           & \checkmark \\ 
  Res2Net152                       & 2                   & \checkmark &                           & \checkmark \\ 
  ERes2Net152                      & 3                   & \checkmark &                           & \checkmark \\ 
  \multirow{2}{*}{ResNet293}       & 4a                  & \checkmark &                           & \checkmark \\ 
                                   & 4b                  &                           & \checkmark & \checkmark \\ 
  \multirow{2}{*}{Res2Net293}      & 5a                  & \checkmark &                           & \checkmark \\ 
                                   & 5b                  &                           & \checkmark & \checkmark \\ 
  \multirow{2}{*}{ERes2Net293}     & 6a                  & \checkmark &                           & \checkmark \\ 
                                   & 6b                  &                           & \checkmark & \checkmark \\ 
  \multirow{2}{*}{Simam-ResNet293} & 7a                  & \checkmark &                           & \checkmark \\ 
                                   & 7b                  &                           & \checkmark & \checkmark \\ \hline
  \end{tabular}}
  \end{table}

\subsection{Training protocol}

Our speaker verification system was implemented with WeSpeaker toolkit \cite{wang2022wespeaker}. All our single models were trained on Nvidia A100 and V100 GPUs. 
Our training protocol encompassed two distinct step.
\begin{table*}[!htbp]
  \centering
  \caption{Performance of models. Results already include AS-Norm and calibration with QMFs.}
  \label{tab:perf}
  \begin{tabular}{cccccccccc}
  \hline
  \multirow{2}{*}{ID} & \multirow{2}{*}{Network}                           & \multicolumn{2}{c}{VoxSRC2023 val}                             & \multicolumn{2}{c}{VoxCeleb1-O}                                  & \multicolumn{2}{c}{VoxCeleb1-E}                                  & \multicolumn{2}{c}{VoxCeleb1-H}                                  \\ \cline{3-10}
     &                                   & EER                          & minDCF                       & EER                          & minDCF                       & EER                          & minDCF                       & EER                          & minDCF                       \\ \hline
  1  & ResNet152                         & 2.827 &  0.147 &  0.420 &  0.025 &  0.578 &  0.032 &  1.038 &  0.057 \\
  2  & Res2Net152                        &  3.163 &  0.170 &  0.532 &  0.028 &  0.665 &  0.040 &  1.187 &  0.068 \\
  3  & ERes2Net152                       &  2.999 &  0.153 &  0.415 &  0.026 &  0.601 &  0.034 &  1.106 &  0.060 \\
  4a &                                   &  2.913 &  0.156 &  0.362 &  0.023 &  0.553 &  0.033 &  1.058 &  0.058 \\
  4b & \multirow{-2}{*}{ResNet293}       & 2.929                        & 0.172                        & 0.457                        & 0.027                        & 0.627                        & 0.036                        & 1.154                        & 0.066                        \\
  5a &                                   &  2.772 &  0.146 &  0.362 &  0.024 &  \textbf{0.543} &  \textbf{0.031} &  \textbf{0.990} &  \textbf{0.054} \\
  5b & \multirow{-2}{*}{Res2Net293}      &  \textbf{2.706} &  0.144 &  0.425 &  0.026 &  0.556 &  0.033 &  1.011 &  0.055 \\
  6a &                                   &  2.772 &  0.149 &  \textbf{0.298} &  \textbf{0.019} &  0.559 & \textbf{0.031} &  1.029 &  0.056 \\
  6b & \multirow{-2}{*}{ERes2Net293}     &  2.847 &  \textbf{0.143} &  0.351 &  0.020 &  0.545 &  0.032 &  1.020 &  0.055 \\
  7a &                                   &  2.948 &  0.165 &  0.372 &  0.024 &  0.608 &  0.037 &  1.199 &  0.067 \\
  7b & \multirow{-2}{*}{Simam-ResNet293} & 2.863                        & 0.177                        & 0.415                        & 0.022                        & 0.629                        & 0.038                        & 1.154                        & 0.065                        \\ \hline
    & Fusion                            & \textbf{2.136}                        & \textbf{0.1260}                        &          -                    &       -                       &                    -         &                      -       & -                              &     -                        \\ 
    & Fusion (test set)                           & \textbf{1.980}                        & \textbf{0.1066}                        &          -                    &       -                       &                    -         &                      -       & -                              &     -                        \\ \hline
     &                                   &                              &                              &                              &                              &                              &                              &                              &                             
  \end{tabular}
  \end{table*}
\subsubsection{Step 1: initial training}

For all models, SGD is used as the optimizer, initialized with a learning rate of 0.1 and decayed to 1e-5 at the end. The learning rate decreased exponentially with a ratio of 1e-4.
The training batch size for models with a depth of 152 was 32, and for models with a depth of 293 was adjusted to 16. 
All models were trained for a total of 150 epochs to ensure model convergence. 

\subsubsection{Step 2: large margin fine-tuning}

Large margin fine-tuning \cite{thienpondt2021idlab} has been widely used for further increasing discriminative ability of speaker embeddings. Some changes were made that speed perturbation and Inter-TopK loss were removed, when doing large margin fine-tuning. 
For all models trained on AMM+K-subcenter+Inter-TopK or Sphereface2, AAM+K-subcenter was the only loss function at the step 2. The margin was set to 0.5 from 0.2.
The length 2s of training segments in step 1 was adjusted to 6s as well.

\subsection{Score procedure and fusion}

Cosine distance was used as the score metric.
Adaptive score normalization \cite{cumani2011comparison} was also applied after score computation, where the imposter cohort size was set to 300. 
The cohort was estimated from the development set of VoxCeleb2, mirroring the methodology employed by the SJTU system. 

Besides, we constructed 30k trials denoted as \textit{Vox2QmfsDev} from the development set for score calibration,  following the strategy \cite{thienpondt2021idlab}. 
We utilized 6 QMF values the same in the ID R\&D system \cite{makarov2022id}: 
\begin{itemize}
  \item \textit{a}) speech length of the enrollment utterance;
  \item \textit{b}) speech length of the test utterance;
  \item \textit{c}) logarithm of the sum of \textit{a} and \textit{b};
  \item \textit{d}) logarithm of the sum of test and enrollment utterance lengths;
  \item \textit{e}) SNR of the test utterance;
  \item \textit{f}) SNR of the enrollment utterance;
\end{itemize}
The final calibrated fusion scores was calculated as
\begin{equation}
  S^{'}=v_0\cdot \mathbf{W}^T\mathbf{S}+\mathbf{V}^T\mathbf{Q}+b
\end{equation}
where $v_0$, $b$ and $\mathbf{V}\in \mathbb{R}^{6\times 1}$ are learnable weights trained on \textit{Vox2QmfsDev}, $\mathbf{W} \in \mathbb{R}^{n\times 1}$ is fixed weight, and $\mathbf{S} \in \mathbb{R}^{n\times 1}$ and $\mathbf{Q} \in \mathbb{R}^{6\times 1}$ are normalized score matrix and QMF values, respectively.

\subsection{Evaluation metric}
For track 1, the speaker verification task, there are two evaluation metrics as followed:
\begin{itemize}
  \item Equal Error Rate (\textit{EER}): the error rate when False Acceptance (FA) and False Rejection (FR) error rates are equal.
  \item minimum detection cost function {\textit{minDCF}}: the cost considering that achieving a low false positive rate is more important than achieving a low false negative rate. The following parameters were used to compute the cost: $C_miss=1$, $C_{FA}=1$, and $P_{Target}=0.05$
\end{itemize}
\section{Results}

Table~\ref{tab:perf} shows the performance of single models and the fusion system on VoxSRC2023 val, VoxCeleb1-O, VoxCeleb-E and VoxCeleb-H. It can be seen that for single model, Res2Net293 and ERes2Net293 achieved the best results, with their respective performance really close to each other. For example, Res2Net293 trained with Sphereface2 loss (ID 5b) obtained the lowest EER of 2.706\% on VoxSRC2023 validation set. It is observed that the performance of ResNet293 and Simam-ResNet293 fell short of initial expectations, which seemed to be attributed to the inadequacy of the training data volume, potentially leading to overfitting.
Score fusion yielded a substantial enhancement in system performance, resulting in a remarkable reduction of EER to 2.136\%, coupled with a decrease in minDCF to 0.1260 on the validation set. Our final submission achieved EER of 1.980\% and minDCF of 0.1066, underscoring the effectiveness of our fusion strategy.

\section{Conclusions}

In this report, we make a detailed description of our solution for Track 1 of VoxSRC 2023. We employed ResNet-based variants with different depths and loss functions. It is suggested that score fusion of these variants plays a significant role for speaker verification, which brings impressive performance improvement.



\bibliographystyle{IEEEtran}
\bibliography{mybib}

\end{document}